\documentclass[a4paper,12pt,oneside]{article}
\usepackage[cp1251]{inputenc}
\usepackage[english]{babel}
\usepackage{amsmath}
\usepackage{amssymb}
\usepackage{epsf}
\usepackage{graphicx}
\usepackage[usenames]{color}
\usepackage{color}

\usepackage{tocenter}
\FromMargins[f]{15mm}{15mm}{20mm}{20mm}

\newcommand{\rot}{\mathop{\mathrm{rot}}\nolimits}
\newcommand{\grad}{\mathop{\mathrm{grad}}\nolimits}

\renewcommand{\Im}{\mathop{\mathrm{Im}}\nolimits}

\title{Refraction and Diffraction of Waves in Electromagnetic (Photonic) Crystals Formed by Anisotropically Scattering Elements}
\author{V.G. Baryshevsky and E.A. Gurnevich\\
Research Institute for Nuclear Problems, Belarusian State
University, \\ 11 Bobruiskaya Str., Minsk 220050, Belarus;
\\ e-mail: bar$@$inp.bsu.by; v\_baryshevsky$@$yahoo.com}

\date{}

\begin{document}
\maketitle
\renewcommand{\figurename}{Figure}
\renewcommand{\refname}{}

\begin{abstract}
Refraction and diffraction of waves in natural crystals  and
artificial crystals formed by  anisotropically scattering centers
are considered. A detailed study of the electromagnetic wave
refraction in a two-dimensional photonic crystal formed by
parallel threads is given by way of example. The expression is
derived for the effective amplitude of wave scattering by a thread
(in a crystal) for the case when scattering by a single thread in
a vacuum is anisotropic. It is established that for a wave with
orthogonal polarization, unlike a wave with parallel polarization,
the index of refraction in  crystals built from metallic threads
can be greater than unity, and Vavilov-Chrernkov radiation becomes
possible in them. The set of equations describing the dynamical
diffraction of waves in  crystals is derived for the case when
scattering by a single center in a vacuum is anisotropic.

Because a most general approach is applied to the description of
the scattering process, the results thus obtained are valid for a
wide range of cases without being restricted to either
electromagnetic waves or crystals built from threads.
 \end{abstract}

\maketitle

\section*{Introduction}

Creation of metamaterials has recently become an area of vigorous
research worldwide. The so-called electromagnetic (photonic)
crystals built from, e.g.,  metallic split rings or parallel
metallic threads \cite{metamaterials}, including threads with
dimensions within the nanometer range \cite{Nanorods} are actively
being studied.
Such photonic crystals can be used, and are already being used,
for solving various tasks, particularly in antenna microwave
technology \cite{metamaterials}. In addition, crystals built from
periodically strained parallel metallic threads can serve as
resonators in volume free electron lasers (VFEL)
\cite{Baryshevsky, FirstGridExp,FirstGridExp1}.

The interaction of electromagnetic waves with photonic crystals is
accompanied by the phenomena of refraction and diffraction.  As is
known, the refractive index of the medium formed by randomly
distributed scatterers is related to the amplitude of scattering
by a single center as follows \cite{Goldberger}
\begin{equation}
  n^2 = 1 + \frac{4\pi\rho}{k^2}A(0),
  \label{eq:n2_3Dchaotic}
\end{equation}
where $\rho$ is the density of scatterers, $A(0)$ is the amplitude
of forward scattering.

However, it is shown in \cite{Baryshevsky1966, BVG95Nuclearoptics}
that if scatterers  are located periodically (e.g. crystal), then
for a correct  description of the refraction process, one should
use a modified expression for $n^2$
\begin{equation}
  n^2 = 1 + \frac{4\pi}{k^2\Omega_3}\frac{A}{1+ikA},
  \label{eq:n2_3Dcrystal}
\end{equation}
where $\Omega_3$ is the volume of the unit cell of the crystal.
 If scattering by a single centers is elastic, formula
 \eqref{eq:n2_3Dcrystal}, in contrast to \eqref{eq:n2_3Dchaotic},
 leads to a physically correct result: the imaginary part of the
 refractive index equals zero.
 Note that  \eqref{eq:n2_3Dcrystal} is derived under the assumption that scattering
 by a single center is isotropic, i.e., the amplitude $A$ is  independent of the scattering angle.
In the present paper, refraction of waves in crystals built from
anisotropically scattering centers is considered. Metallic threads
are the simplest example of such scatterers: scattering by such
threads of electromagnetic waves with the electric-field vector
polarized orthogonally to the thread axis is anisotropic for all
wavelengths \cite{Gurnevich2009, Gurnevich2010}.

This paper is arranged as follows: Section 1 gives a detailed
analysis of a nonplane waves scattering  by a single thread.
Section 2 describes the method used for finding the refractive
index by the example of crystals formed by isotropic scatterers.
Section 3 considers the case of crystals formed by anisotropic
scatterers.

\section{Scattering of Electromagnetic Waves with Orthogonal Polarization
by Thread}

\subsection{Plane Wave}

Before we start to consider scattering of a {\it cylindrical} wave
by a  thread (cylinder), let us recall the case of a plane wave.
Assume, as usual, that the radius $R$ of the thread is much
smaller than its length $L$, and so in analytical treatment, the
thread can be considered infinitely long.

\begin{figure}[htp]
 \centering
 \includegraphics[scale=0.8]{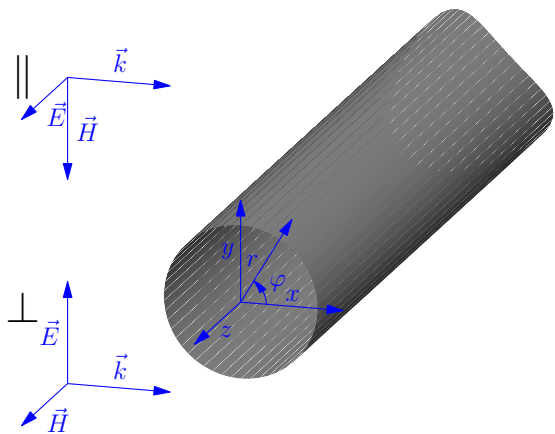}
 \caption{Diffraction of a plane electromagnetic wave by a cylinder.}
 \label{fig:cylinder}
\end{figure}

The solution to the problem of diffraction of a plane
electromagnetic wave by an infinite cylinder can be found in the
form of a series over Bessel and Hankel functions \cite{Nikolsky}.
Let a wave with perpendicular polarization (vector $\vec{E}$ is
perpendicular to the axis of a cylinder) be scattered by a
cylinder placed in a vacuum (Fig.\ref{fig:cylinder}). The axis of
the cylinder coincides with the $z$-axis of the rectangular
coordinate system. Let us also introduce a cylindrical coordinate
system $(r,\varphi, z)$, as is shown in Fig.\ref{fig:cylinder}.

For simplicity, we shall consider the case when the wave vector
$\vec{k}$ is perpendicular to the axis of the cylinder. Then the
field of the scattered wave can be written in the
form~\cite{Nikolsky}:
\begin{equation}
\label{eq:FieldOutsidePerp}
\begin{split}
 \vec{H}_e &= \vec{e}_z \sum\limits_{n=-\infty}^{\infty} i^n c_n^\perp
H_n(kr)e^{-in\varphi}, \\
\vec{E}_e &= \frac{ic}{\omega} \sum\limits_{n=-\infty}^{\infty}
i^n c_n^\perp \left[ \vec{e}_r \frac{-in}{r} H_n(kr) -
\vec{e}_\varphi k H_n'(kr) \right]e^{-in\varphi},\end{split}
\end{equation}
where the coefficients $c_n^\perp$ are calculated by the formula
\begin{equation}
 c_n^\perp = \frac{-J_n(k_2R)J'_n(kR) +
\frac{1}{\sqrt{\varepsilon}}J_n'(k_2R)J_n(kR)}
{J_n(k_2R)H_n^{(1)'}(kR) - \frac{1}{\sqrt{\varepsilon}}
J_n'(k_2R)H_n^{(1)}(kR)}. \label{eq:c_perp}
\end{equation}
Here $J_n$ is the Bessel function of order  $n$, $H_n$ is the
Hankel function of the first kind of order $n$,
$k_2=k\sqrt{\varepsilon}$, $\varepsilon$ is the dielectric
permittivity of the thread (for metals,  $\varepsilon = 1 +
4\pi\sigma i/\omega$, where $\omega$ is the frequency of the
electromagnetic field, $\sigma$ is the conductivity, and the
magnetic permittivity $\mu$ is assumed to be equal to 1), vectors
$\vec{e}_r$, $\vec{e}_\varphi$, $\vec{e}_z$ are the unit vectors
of the cylindrical coordinate system, and the amplitude of the
incident wave is assumed to be equal to  1.

Considering the limits of Bessel and Hankel functions of small
argument, one can find that for a perfectly conducting cylinder,
the  relationships $c_0^\perp \approx -c_{\pm 1}^\perp$ and
$c_0^\perp \gg c^\perp_{n}$ hold for all $|n|>1$ at $kR \ll 1$.
If a cylinder is not a perfect conductor, then the first equality
is violated, but the absolute values of the coefficients
$c_0^\perp$ and $c_{\pm 1}^\perp$ remain comparable.
What is more, if the dielectric permittivity of a cylinder is
$\varepsilon-1\sim 1$, then $c_0^\perp \ll c_{\pm 1}^\perp$.
Thus, at $kR\ll 1$, in 
  \eqref{eq:FieldOutsidePerp}, it
suffices to take account of only the terms $n=0$ and $n=\pm 1$.
Note here that in considering a wave with parallel polarization in
the long-wave limit, $c_0^\parallel \gg c_n^\parallel$ for all
$n\neq 0$, and so in the series for the field, one can take
account of the term $n=0$ alone.

Returning to our problem, let is recall that the two equations
\eqref{eq:FieldOutsidePerp} are related through Maxwell's
equations, and so the second equation can readily be obtained from
the first one by formula $\vec{E}_e = \frac{ic}{\omega}\rot
\vec{H}_e$. With thus eliminated electric field and with due
account of the above remarks concerning the values of the
coefficients $c_n$ (it is further assumed that $kR\ll 1$), one can
write the scattered wave  in the form:
\begin{equation}
 \vec{H}_e = \vec{e}_z \Psi_{sc} = \vec{e}_z \left\{ c_0H_0(kr) +
2ic_1H_1(kr)\cos\varphi \right\},
\end{equation}
where the superscript $\perp$ on the expansion coefficient is
dropped.

If a plane wave $\vec{H}_0 = \Psi_0\vec{e}_z = e^{ikx}\vec{e}_z$,
is scattered by a cylinder placed at the origin of coordinates,
then the total wave field is expresses as a sum of the field of
the incident and scattered waves, i.e.,
\begin{equation}
 \Psi = \Psi_0 + \Psi_{sc} = e^{ikx} +  c_0H_0(kr) + 2ic_1H_1(kr)\cos\varphi.
\label{eq:psi_sc_pm1}
\end{equation}

Using the asymptotic expression for Hankel functions of large
argument  and the integral representation of Hankel functions, the
above expression  can be presented in the following form, provided
$kr \gg 1$:
\begin{equation}
 \Psi = e^{ikx} +
A(\varphi)\int\limits_{-\infty}^{\infty}\frac{e^{ik\sqrt{r^2+z^2}}}{\sqrt{r^2+z^2}}
\mathrm{d}z, \label{eq:asymptotic}
\end{equation}
where $A(\varphi) = -\frac{i}{\pi}(c_0 + 2c_1\cos\varphi) = A_0 +
A_1\cos\varphi$.

Similarly to a three-dimensional case, by $A(\varphi)$ one should
understand the amplitude of scattering of an electromagnetic wave
by a thread at an angle
 $\varphi$ \cite{landau77quant}.

It should be noted that in contrast to a diverging spherical wave
that characterizes scattering in a three-dimensional case, in a
two-dimensional case, a diverging cylindrical wave is formed.

 In
view of \eqref{eq:psi_sc_pm1}, one can readily write the
expression for the field in the case when the axis of a cylinder
lies not in the origin of coordinates, but at point $\vec{r}_1$
\begin{equation}
  \Psi = \Psi_0 + \Psi_{sc} = e^{i\vec{k}\vec{r}} + e^{i\vec{k}\vec{r}_1}\cdot A_0 i\pi
  H_0(k|\vec{r}-\vec{r}_1|) - e^{i\vec{k}\vec{r}_1}\cdot A_1 \pi H_1(k|\vec{r}-\vec{r}_1|)
  \cos(\vec{k},\vec{r}-\vec{r}_1).
  \label{eq:eq1}
\end{equation}

\subsection{Cylindrical wave}

Let now a cylindrical wave $\Psi_0 = H_0(kr)$, diverging from the
origin of coordinates,  be incident onto this scatterer, which is
placed at point $\vec{r}_1$.
To analyze the scattering process in this case, one can decompose
a cylindrical wave into elementary plane waves, which are
scattered according to a well-known law \eqref{eq:eq1}.
For the Hankel function, in particular, one can use the following
representation:
\begin{equation}
  H_0(kr) = \frac{1}{\pi}\int\limits_{-\infty}^{\infty}\frac{e^{i\vec{k}\vec{r}}}{\sqrt{k^2-k_y^2}}dk_y =
\frac{1}{\pi}\int\limits_{-\infty}^{\infty}\frac{e^{i\vec{k}\vec{r}}}{\sqrt{k^2-k_x^2}}dk_x.
  \label{eq:H0}
\end{equation}

For the purposes of this paper, we do not need to know the exact
decomposition, suffice it to know that the wave $\Psi_0=H_0(kr)$
can be presented as a sum of plane waves as follows: $H_0(kr)=\int
f(k,k_y)e^{i\vec{k}\vec{r}}dk_y$, or $H_0(kr)=\int
f(k,k_x)e^{i\vec{k}\vec{r}}dk_x$, where $k\equiv |\vec{k}| =
const$ and $f(k,k_y)$ is a certain function.

Thus, by decomposing the initial wave and in view of
\eqref{eq:eq1}, one can immediately write the expression for a
scattered wave \footnote{In \eqref{eq:scattered1},
$e^{i\vec{k}\vec{r}_1}\equiv e^{i\sqrt{k^2-k_y^2}x_1 + ik_yy_1}$;
at $k_y>k$, this quantity is a damped plane wave  whose wave
number is $k_y$, but not $k$. At large values of $k_y$ ($k_yR\gg
1$), the coefficients $c_n$ ($n>1$) for such a wave are comparable
with $c_0$ and $c_1$ (see~\eqref{eq:c_perp}). And so, in
integrating over the domain of large values of $k_y$, it would be
desirable to add in \eqref{eq:scattered1} the terms of the form
\begin{equation*}
  H_n(k_y|\vec{r}-\vec{r}_1|)\times \int\limits_{|k_yR|\gg 1}\frac{e^{-\sqrt{k_y^2-k^2}x_1}
e^{ik_yy_1}}{i\sqrt{k_y^2-k^2}} c_n(k_y) \cos(n\alpha) dk_y,
\end{equation*}
where  $\alpha$ is the angle between vectors  $\vec{k}$ and
$\vec{r}-\vec{r}_1$. Simple estimates, however, show that the
absolute values of these integrals are small  ($\sim e^{-k_y^{min}
x_1}\ll 1$), and so the corresponding additions can be neglected.}
\begin{equation}
  \Psi_{sc} = \int\limits_{-\infty}^{\infty}f(k,k_y)e^{i\vec{k}\vec{r}_1}A_0i\pi H_0(k|\vec{r}-\vec{r}_1|)dk_y
  - \int\limits_{-\infty}^{\infty}f(k,k_y)e^{i\vec{k}\vec{r}_1}A_1\pi H_1(k|\vec{r}-\vec{r}_1|)
  \cos(\vec{k},\vec{r}-\vec{r}_1)  dk_y.
  \label{eq:scattered1}
\end{equation}
Because $A_{0,1}$ and  $H_{0,1}(k|\vec{r}-\vec{r}_1|))$ are only
dependent on the absolute value of the wave vector $k$ and
independent of $k_y$, they can be removed from the integration
sign.
Then, according to   \eqref{eq:H0}, the first integral in
\eqref{eq:scattered1} equals $H_0(kr_1)$, and
\eqref{eq:scattered1} can be rewritten in the form:
\begin{equation}
  \Psi_{sc} = H_0(kr_1)\cdot A_0i\pi H_0(k|\vec{r}-\vec{r}_1|) - A_1\pi H_1(k|\vec{r}-\vec{r}_1|)\cdot I_1,
  \label{eq:scattered2}
\end{equation}
where
\begin{equation}
  I_1 = \int\limits_{-\infty}^{\infty}f(k,k_y)e^{i\vec{k}\vec{r}_1}\cos(\vec{k},\vec{r}-\vec{r}_1)  dk_y.
\end{equation}
To evaluate the integral  $I_1$, let us represent the cosine of
the angle  between  vectors  $\vec{k}$ and $\vec{r}-\vec{r}_1$ in
the form:
$\cos(\vec{k},\vec{r}-\vec{r}_1)=\frac{\vec{k}(\vec{r}-\vec{r}_1)}{k|\vec{r}-\vec{r}_1|}$.
Then we have
\begin{multline}
  I_1 = \frac{\vec{r}-\vec{r}_1}{k|\vec{r}-\vec{r}_1|}\cdot\frac{1}{\pi}\int\limits_{-\infty}^{\infty}
  \vec{k} f(k,k_y)e^{i\vec{k}\vec{r}_1}dk_y =\\= -i\frac{\vec{r}-\vec{r}_1}{k|\vec{r}-\vec{r}_1|}\cdot
  \grad_1\int\limits_{-\infty}^{\infty}
  f(k,k_y)e^{i\vec{k}\vec{r}_1}dk_y  = -i\frac{\vec{r}-\vec{r}_1}{k|\vec{r}-\vec{r}_1|}\cdot
  \grad_1H_0(kr_1),
  \label{eq:integral}
\end{multline}
where the subscript ``1'' on the gradient symbol means that
differentiation is performed with respect to the coordinates of
the point $\vec{r_1}$. The differentiation yields $\grad_1
H_0(kr_1) = -H_1(kr_1)\frac{k\vec{r}_1}{r_1}$, and then
\begin{equation}
  I_1 = iH_1(kr_1) \frac{\vec{r}_1 (\vec{r}-\vec{r_1})}{r_1 |\vec{r}-\vec{r_1}|} =
  iH_1(kr_1)\cos(\vec{r}_1, \vec{r}-\vec{r}_1).
  \label{eq:I1}
\end{equation}
The above yields that the total wave field  for scattering of a
cylindrical wave  has the form:
\begin{equation}
  \Psi = H_0(kr) + A_0i\pi H_0(kr_1)\cdot H_0(k|\vec{r}-\vec{r}_1|) - A_1\pi iH_1(kr_1)\cdot
  H_1(k|\vec{r}-\vec{r}_1|)\cos(\vec{r}_1, \vec{r}-\vec{r}_1).
  \label{eq:scatteredH_0}
\end{equation}
At large distances  from the origin of coordinates
 ($kr_1 \gg 1$), one can use the serial expansion $iH_1(z)\approx
H_0(z)\left(1 + \frac{i}{4z}\right)$. Substitution of this
expression into \eqref{eq:scatteredH_0} gives
\begin{equation}
\Psi = H_0(kr) + A_0i\pi H_0(kr_1)\cdot H_0(k|\vec{r}-\vec{r}_1|)
- A_1\pi H_0(kr_1)\left( 1 + \frac{i}{4kr_1}\right) \cdot
H_1(k|\vec{r}-\vec{r}_1|)\cos(\vec{r}_1, \vec{r}-\vec{r}_1).
  \label{eq:scatteredApprox}
\end{equation}
Comparing this equation with \eqref{eq:eq1}, one can conclude that
with increasing distance between the origin of coordinates (the
central point  of the diverging cylindrical wave) and the
scatterer, the difference between the amplitudes of scattering of
cylindrical and plane waves by a thread  diminishes, as might be
expected.

Let us briefly consider the case when the initial cylindrical wave
has the form $\Psi_0 = H_1(kr)\cos\varphi$.
 By reasoning along the same lines, we
obtain similar formulas, except for the expression for the
integral $I_1$. Now it equals
\begin{equation}
  I_1' = -i\frac{\vec{r}-\vec{r}_1}{k|\vec{r}-\vec{r}_1|}\cdot
  \grad_1 H_1(kr_1)\cos\varphi_1.
  \label{eq:I2}
\end{equation}
Now differentiation yields the following expression
\begin{equation}
  \grad_1 H_1(kr_1)\cos\varphi_1 = \left\{kH_0(kr_1) - \frac{2}{r_1}H_1(kr_1)\right\}
  \cos\varphi_1\cdot\frac{\vec{r}_1}{r_1} + \frac{1}{r_1}H_1(kr_1)\vec{e}_x,
\end{equation}
where $\vec{e}_x$ is the unit vector of the $x$-axis. The
expression for a scattered wave will now have the form:
\begin{multline}
  \Psi = \Psi_0 + \Psi_{sc} = H_1(kr)\cos\varphi +
  \left\{A_0H_1(kr_1)\cos\varphi_1\right\}i\pi H_0(k|\vec{r}-\vec{r}_1|) -\\-
  \left\{A_1\cos\varphi_1 \left( -iH_0(kr_1) + \frac{2i}{kr_1}H_1(kr_1) \right)\right\}\pi
  H_1(k|\vec{r}-\vec{r}_1|) \cos(\vec{r}_1,\vec{r}-\vec{r}_1) -\\-
  \left\{\frac{-iA_1}{kr_1}H_1(kr_1)\right\} \pi H_1(k|\vec{r}-\vec{r}_1|)\cos(\vec{r}-\vec{r}_1,\vec{e}_x).
  \label{eq:scatteredH1cos}
\end{multline}
A particular case when the scatterer is placed on the $y$-axis,
i.e.,  $\varphi_1=\pm\frac{\pi}{2}$,  seems to be of interest.
Here we have $\cos\varphi_1=0$, and the expression for a scattered
wave simplifies appreciably:
\begin{equation}
  \Psi_{sc} = \frac{i\pi A_1}{kr_1}H_1(kr_1) H_1(k|\vec{r}-\vec{r}_1|)\cos(\vec{r}-\vec{r}_1,\vec{e}_x).
  \label{eq:scatteredH1cospina2}
\end{equation}
It should be noticed that here  the scattered wave does not vanish
despite the zero amplitude of the incident wave at the scatterer
location: $\Psi_0(\vec{r}_1)=H_1(kr_1)\cos \varphi_1=0$. Though
there is not any paradox because at $\varphi_1=\pm \frac{\pi}{2}$,
only the magnetic-field vector equals zero,
$\vec{H}_0=\Psi_0\vec{e}_z$, while the electric-field vector has
the form:  $\vec{E}_0=\frac{ic}{\omega}\rot
\vec{H}_0=-\frac{ic}{\omega r_1}H_1(kr_1)\vec{e}_y$, i.e., is
nonzero. Thus, when $\varphi_1=\pm\frac{\pi}{2}$, only the
electrical component of the initial wave is scattered by the
thread.

\section{Refraction of Waves in  Photonic Crystals Formed by Isotropically Scattering Elements}

\begin{figure}[ht]
  \begin{center}
    \includegraphics[scale=0.5]{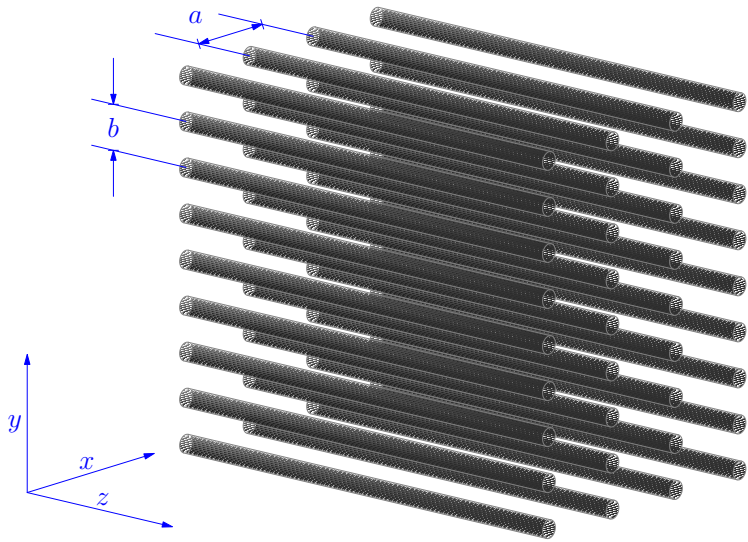}
  \end{center}
  \caption{Crystal made of parallel metallic threads}
  \label{fig:crystal}
\end{figure}

Let us consider an infinite, two-dimensional  crystal composed of
periodically arranged scatterers.  A well-known example of such a
crystal is a photonic crystal built from parallel metallic threads
\cite{Baryshevsky}, Fig.~\ref{fig:crystal}.

For definiteness (but without loss of generality!), we shall
consider such a crystal. Let the  coordinates of the threads in
the lattice be $(x_{mn}, y_{mn}) = (ma, nb)$, where $a$ and $b$
are the lattice spacings, $m$ and $n$ are the integers; the axes
of the treads are parallel to the $z$-axis of the rectangular
coordinate system.

One should distinguish between the cases when vector $\vec{E}$ of
the incident wave is polarized parallel to the threads and when it
is polarized perpendicular to them.
In the first case, scattering at $kR\ll 1$ is isotropic and
$A(\varphi)=A_0$, while in the second case, scattering is
anisotropic even at $kR\ll 1$, and the amplitude has the above
described angular dependence: $A=A_0+A_1\cos\varphi$.
Certainly, when $kR\gtrsim1$, in expression \eqref{eq:psi_sc_pm1}
one needs to retain more terms in the expansion
\eqref{eq:FieldOutsidePerp}, and the angular dependence of the
amplitude can take a more complicated form.

In this section, we shall assume that scattering by the centers is
isotropic with the amplitude $A_0$.
Let us suppose that an electromagnetic wave $\sim e^{iqx}$
propagates in a crystal in the positive direction of the $x$-axis.
\footnote{The wave vectors $\vec{k}$ and $\vec{q}$ are assumed to
be perpendicular to the $z$-axis. The case of  arbitrary incidence
of a  wave onto the $z$-axis will be considered individually.}
This wave results from the summation of diverging cylindrical
waves $i\pi H_0(k|\vec{r}-\vec{r}_{mn}|)$ radiated by all the
threads. The amplitude of these waves has the form
$\Phi=\Phi_0e^{iqx_{mn}}$, where $\Phi_0$ is independent of the
position of the thread in the crystal.
Using the method  described in \cite{Belov1}, let us find the
relation between the wave numbers $k$ and $q$. Let us assume that
the local field at the location of each thread is a sum of all
waves coming from all other threads.
Particularly, the local filed acting on the thread placed at the
origin of coordinates ($m=n=0$) can be written as follows:
\begin{equation}
  \Psi_{loc} = \Phi_0i\pi\sum\limits_{(m,n)\neq (0,0)} e^{iqx_{mn}}H_0(kr_{m,n}) =
  \Phi_0i\pi\sum\limits_{(m,n)\neq (0,0)}e^{iqam}H_0\left( k\sqrt{(am)^2 + (bn)^2} \right) .
  \label{eq:sum1}
\end{equation}

Since scattering is isotropic, all these waves are scattered by
this thread with a known amplitude, equal to the amplitude $A_0$
of scattering of a plane wave by the thread, producing a diverging
cylindrical wave with the amplitude $\Phi_0$:
\begin{equation*}
  \Phi_0 = A_0\Psi_{loc} = A_0i\pi \Phi_0\sum\limits_{(m,n)\neq (0,0)}e^{iqam}H_0\left(
  k\sqrt{(am)^2 + (bn)^2} \right).
\end{equation*}
We thus come to the dispersion equation  for finding $q$
\begin{equation}
  \frac{1}{A_0i\pi} = \sum\limits_{(m,n)\neq (0,0)}e^{iqam}H_0\left( k\sqrt{(am)^2 + (bn)^2} \right)i = S(k,q).
  \label{eq:disp_isotrop}
\end{equation}

The sum in  \eqref{eq:disp_isotrop} is taken as described in
\cite{Belov1}, and here we only give the resulting expression:
\begin{equation}
  S(k,q) = -1 -i\cdot\left\{ \frac{2}{\pi}\left(\log\frac{kb}{4\pi} + C\right) +
  \frac{2}{kb}\frac{\sin ka}{\cos ka - \cos qa} + \frac{2}{b}
  \sum\limits_{n\neq0} \frac{1}{k_x^{(n)}}\frac{\sin k_x^{(n)}a}{\cos k_x^{(n)}a - \cos qa} -
  \frac{b}{2\pi |n|}  \right\},
  \label{eq:sum}
\end{equation}
where $k_x^{(n)} = i\sqrt{(2\pi n/b)^2 - k^2}$, $C\approx0.5772$
is the Euler constant. Let us assume that the scattering amplitude
is sufficiently small: $|A_0(k)| \ll 1$ and the refractive index
is close to unity: $|n-1|=|q/k -1| \ll 1$.

One can easily demonstrate that in this case, the second term
between the braces is a dominating term, and far from the
diffraction conditions ($ka\neq\pi n$) it can be presented in the
form:
\begin{equation}
  \frac{2}{kb}\frac{\sin ka}{\cos ka - \cos qa} \simeq \frac{4}{k^2 ba}\cdot\frac{1}{n^2 - 1}
  = \frac{4}{k^2\Omega_2}\cdot\frac{1}{n^2 - 1},
  \label{eq:secondterm}
\end{equation}
where $\Omega_2 = ab$ is the crystal unit cell area. We thus come
to the following dispersion equation:
\begin{equation*}
  \frac{1}{A_0i\pi} \approx -1 - \frac{4i}{k^2\Omega_2}\cdot \frac{1}{n^2-1},
\end{equation*}
from this we get
\begin{equation}
  n^2 = 1 + \frac{4\pi}{k^2 \Omega_2}\cdot \frac{A_0}{1 + i\pi A_0}.
  \label{eq:rindex_iso}
\end{equation}
%

At this point, it should be mentioned that the same result was
obtained in \cite{Baryshevsky} in a different way. In the case of
elastic scattering (e.g. perfectly conducting threads), formula
\eqref{eq:rindex_iso} gives a real value of the refractive index.
This is easily verified by the optical theorem. It will be
recalled that the optical theorem relates the imaginary part of
the forward scattering amplitude to the total scattering cross
section: $\Im A(0) = \frac{k\sigma}{4\pi}$. For a two-dimensional
case, the differential scattering cross section is related to the
scattering amplitude as follows:
$\frac{d\sigma}{d\varphi}=\frac{2\pi}{k}|A(\varphi)|^2$.

One can arrive at the same result using a slightly different
method: start with the consideration of  scattering of a plane
wave by a one-dimensional grating built from threads and then
proceed to the case of infinite crystals.
As this method is more convenient for the purposes of the analysis
given in the following section, let us also describe it here.

\begin{figure}[ht]
  \begin{center}
    \includegraphics{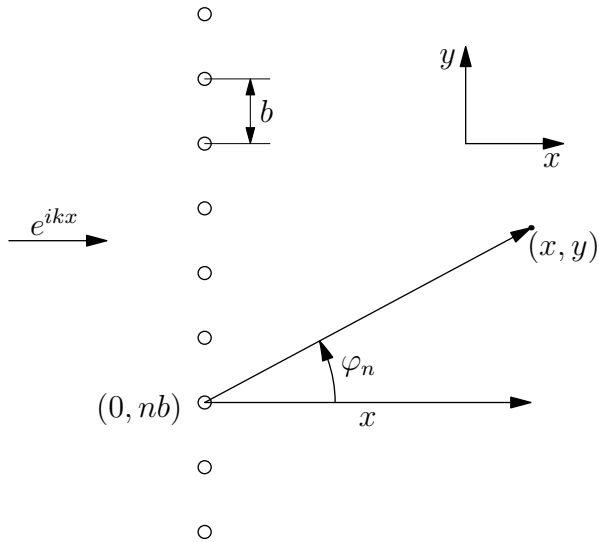}
  \end{center}
  \caption{Diffraction of a plane wave by a one-dimensional array of scatterers}
  \label{fig:odnomer}
\end{figure}

Let a plane wave be scattered by a one-dimensional array of
scatterers, as is shown in Fig. ~\ref{fig:odnomer}.
The scattered wave in this case is a sum of cylindrical waves of
the same amplitude $F$, which are radiated  by each thread
\begin{equation*}
  \Psi_{sc}(x,y) = F\sum\limits_{n=-\infty}^{\infty}i\pi H_0(k\sqrt{x^2 + (y-bn)^2}).
\end{equation*}
But the amplitude $F$ of the  plane-wave scattering by  a thread
in the presence of other scatterers differs from the amplitude
$A_0$ of scattering by a single thread. With the help of the
general methods used for describing multiple scattering
\cite{Goldberger}, the amplitude $F$ can be expressed in terms of
$A_0$ as follows:
 \begin{equation}
  F = A_0 + A_0F\sum\limits_{n\neq 0}i\pi H_0(kb|n|).
  \label{eq:amp0_sheet}
\end{equation}

The physical meaning of equations \eqref{eq:amp0_sheet} is
obvious: there are two waves scattered by each thread
(particularly, by the thread placed at the origin of coordinates):
the plane wave $e^{ikx}$ and the wave $\Psi'=F\sum\limits_{n\neq
0}i\pi H_0(kb|n|)$ scattered by all other threads with the
amplitude  $A_0$. The sum \eqref{eq:amp0_sheet} is taken in
\cite{Belov1}
\begin{equation}
  S_1 = \sum\limits_{n\neq 0}H_0(kb|n|) = \frac{2}{kb} -1 -i\frac{2}{\pi} \left( C + \log \frac{kb}{4\pi}
  \right) - \frac{2i}{b}\sum\limits_{n\neq 0}
  \left( \frac{1}{\sqrt{(2\pi n/b)^2 - k^2}} - \frac{b}{2\pi |n|} \right).
  \label{eq:sum_sheet}
\end{equation}
It is still assumed that $A_0\ll 1$, then using
\eqref{eq:amp0_sheet} and  \eqref{eq:sum_sheet}, one can derive
the following expression\footnote{To simplify the derived
expressions, in writing \eqref{eq:amp0_s_val} and
\eqref{eq:plane_wave}, we have made the inessential assumption
that $kb<2\pi$. Quite simple, but cumbersome calculations
(subsequent consideration of  the cases $2\pi<kb<4\pi$,
$4\pi<kb<6\pi$, and so on) show that this assumption has no
influence on the final result, which remains valid for large
values of $k$ (far from the diffraction conditions) too.} for the
effective scattering amplitude:
\begin{equation}
  F = \frac{A_0}{1 - i\pi S_1 A_0} \simeq \frac{A_0}{1 + i\pi A_0 - i\frac{2\pi}{kb}A_0}.
  \label{eq:amp0_s_val}
\end{equation}
Thus the wave field resulting from scattering of a plane wave by a
one-dimensional grating  has the form: \cite{Baryshevsky,
Gurnevich2009}
\begin{equation}
  \Psi = e^{ikx} + F\sum\limits_{n=-\infty}^{\infty}i\pi H_0(k\sqrt{x^2 + (y-bn)^2}) \approx
  e^{ikx} + \frac{2i\pi}{kb}F e^{ik|x|},
  \label{eq:plane_wave}
\end{equation}
where $F$ is defined from \eqref{eq:amp0_s_val}, and  is a sum of
the wave fields of the plane wave incident onto the grating and
the wave scattered by the grating. The scattered wave propagates
in the positive direction of the $x$-axis at $x>0$, and in the
negative direction at $x<0$, its amplitude in both cases being
$\frac{2i\pi}{kb}F$. This quantity can be treated as the
"amplitude of scattering"  of a plane wave by a grating built from
threads.
%

Now let us consider the  propagation of waves in a crystal formed
by an infinite periodic  system of plane gratings regularly spaced
at an interval $a$ and composed of threads. Use the same method as
before to derive the dispersion equation. The wave incident onto
the grating placed at the origin of coordinates is a sum of waves
scattered by all other gratings. Assuming that the amplitude of
these waves has the form $\Phi = \Phi_0 e^{iqam}$ with $\Phi_0$
being independent of the coordinates due to the periodicity of
crystals, one can write the following expression
\begin{equation*}
  \Phi_0 = \frac{2i\pi}{kb}F \sum\limits_{m\neq 0}\Phi_0 e^{iqam} e^{ika|m|},
\end{equation*}
from which one can derive the dispersion equation  (compare with
\eqref{eq:disp_isotrop})
\begin{equation}
  1 = \frac{2i\pi}{kb}F \sum\limits_{m\neq 0}e^{iqam} e^{ika|m|}.
  \label{eq:disp1}
\end{equation}
The sum in  \eqref{eq:disp1} can easily be calculated using the
 geometric progression sum formula, and is equal to
\begin{equation*}
  \sum\limits_{m\neq 0}e^{iqam} e^{ika|m|} = -1 -i \frac{\sin ka}{\cos ka - \cos qa} \simeq -1 -
  \frac{2i}{ka}\cdot \frac{1}{n^2 -1}.
\end{equation*}
By substituting the obtained value into equation \eqref{eq:disp1}
and using  \eqref{eq:amp0_s_val}, we get the already known result
\begin{equation*}
  n^2 = 1 + \frac{4\pi}{k^2\Omega_2}\cdot \frac{A_0}{1 + i\pi A_0}.
\end{equation*}
Note here that similar consideration  of   a three-dimensional
crystal case gives a well-known formula~\eqref{eq:n2_3Dcrystal}.

\section{Refraction of Waves in Crystals. The Case of Anisotropic Scatterers}

Assume now that the scattering amplitude has the form
$A=A_0+A_1\cos\varphi$, i.e., scattering is anisotropic. Let us
use the approach described in the previous section to consider the
propagation of plane waves in a crystal composed of anisotropic
scatterers.

Let a plane wave  having the amplitude $e^{ikx}$ be incident onto
a one-dimensional grating composed of anisotropically scattering
centers (Fig. \ref{fig:odnomer}).
Knowing how a {\it cylindrical} wave is scattered by such
scatterers (see \eqref{eq:scatteredH_0},
\eqref{eq:scatteredH1cos}), one can determine that the angular
dependence of the amplitude $F(\varphi)$ of scattering by each
thread in the presence of other threads is the same as  that of
the amplitude $A(\varphi)$ of scattering by a single thread
$F(\varphi) = F_0 + F_1\cos \varphi$. The system of equations  for
finding  $F_0$ and  $F_1$ has the form (see
\eqref{eq:scatteredH_0} and \eqref{eq:scatteredH1cospina2}):
\begin{eqnarray}
  F_0 & = &  A_0 + A_0F_0\sum\limits_{m\neq 0} i\pi H_0(kb|m|), \\ \label{eq:eqF0}
  F_1 & = &  A_1 + A_1F_1\sum\limits_{m\neq 0} i\pi H_1(kb|m|) \cdot (kb|m|)^{-1}. \label{eq:eqF1}
\end{eqnarray}
The sum appearing in  equation \eqref{eq:eqF1} can be calculated
 using the known value of the first sum \eqref{eq:sum_sheet} and the following equality for the Hankel function:
\begin{equation*}
  \frac{H_1(kb|n|)}{kb|n|} = -\frac{2i}{\pi k^2b^2n^2} + \frac{1}{k^2}\int\limits_{0}^{k} kH_0(kb|n|) dk.
\end{equation*}
Integration gives:
\begin{equation}
  S_2 = \sum\limits_{m\neq 0} \frac{H_1(kb|m|)}{kb|m|} = \frac{2}{kb} - \frac{1}{2}
  -\frac{i}{\pi}  \left( C - \frac{1}{2} + \log \frac{kb}{4\pi} + \frac{2\pi^2}{3k^2b^2} \right)
  - \frac{2i}{k^2b}\sum\limits_{n\neq 0}
  \left( \frac{2\pi |n|}{b} - \frac{k^2b}{4\pi |n|} - \sqrt{ \frac{4\pi^2n^2}{b^2} - k^2} \right).
  \label{eq:sum_S2}
\end{equation}
Then, as before (see \eqref{eq:amp0_s_val}), we can write the
expressions for the amplitudes $F_0$ and  $F_1$ using equations
\eqref{eq:sum_sheet} and  \eqref{eq:sum_S2}
\begin{eqnarray}
  F_0 & = & \frac{A_0}{1 - i\pi S_1} \simeq \frac{A_0}{1 + i\pi A_0 -i \frac{2\pi}{kb}A_0},
  \label{eq:amp10_s_val}
  \\
  F_1 & = & \frac{A_1}{1 - i\pi S_2} \simeq \frac{A_1}{1 + i\frac{\pi}{2} A_1 -i \frac{2\pi}{kb}A_1}.
  \label{eq:amp1_s_val}
\end{eqnarray}
So the scattered wave field at a point with coordinates   $(x,y)$
has the form:
\begin{equation}
\Psi(x,y) = e^{ikx} + A_0 \sum_{n=-\infty}^{\infty} i\pi
H_0\left(k \sqrt{\left(y-nb\right)^2 + x^2}\right) + A_1
\sum_{n=-\infty}^{\infty} (-\pi)\cdot H_1
\left(k\sqrt{\left(y-nb\right)^2 + x^2} \right) \cos\varphi_n,
\label{eq:psi_sum_1}
\end{equation}
where
\begin{equation}
 \cos\varphi_n = \frac{x}{\sqrt{(y-nb)^2 + x^2}}.
\label{eq:cos_phi_n}
\end{equation}
Summation of these series is given, e.g., in \cite{Gurnevich2009};
finally, for sufficiently large distances $|x|$ from the grating,
we have
\begin{equation}
  \Psi = e^{ikx} + \frac{2i\pi}{kb}  \left( F_0 \pm F_1 \right) e^{ik|x|},
  \label{eq:psi_sum_2}
\end{equation}
where the sign  ``$+$'' refers to the case when  $x>0$ and the
sign ``$-$'', to the case when  $x<0$ (forward and backward
scattering, respectively).

Now let us consider a crystal composed of a number of such
gratings regularly spaced at an interval $a$. The wave $\Psi_1$,
which falls onto the grating placed at the origin of coordinates,
is a sum of plane waves scattered by all other gratings. Let us
assume that the amplitude of these waves is $\Phi = (\Phi_0 \pm
\Phi_1)e^{iqam}$, where $\Phi_0$ and $\Phi_1$ are independent of
the grating number due to the periodicity of the crystal. Then the
wave $\Psi_1$ has the form
\begin{equation}
  \Psi_1 = \frac{2\pi i}{kb}\Phi_0 \sum\limits_{m\neq 0} e^{iqam} e^{ika|m|} +
  \frac{2\pi i}{kb}\Phi_1  \left\{ \sum\limits_{m=-\infty}^{-1}e^{iqam}e^{ika|m|}  -
    \sum\limits_{m=1}^{\infty}e^{iqam}e^{ika|m|}\right\}.
  \label{eq:psi_pad}
\end{equation}
In view of  \eqref{eq:psi_pad} and  \eqref{eq:psi_sum_2}, one can
write the following system of equations for the amplitudes
\begin{eqnarray}
  \Phi_0 & = & F_0 \left\{ \frac{2\pi i}{kb} \Phi_0 S_3 + \frac{2\pi i}{kb} \Phi_1 S_4 \right\},
  \label{eq:phi0}\\
  \Phi_1 & = & F_1 \left\{ \frac{2\pi i}{kb} \Phi_0 S_4 + \frac{2\pi i}{kb} \Phi_1 S_3 \right\},
  \label{eq:phi1}
\end{eqnarray}
where the sums $S_3$ and  $S_4$ are equal to
\begin{eqnarray*}
  S_3 = \sum\limits_{m\neq 0} e^{iqam}e^{ika|m|} = -1 -i \frac{\sin ka}{\cos ka - \cos qa} \simeq -1 -
  \frac{2i}{ka}\cdot \frac{1}{n^2 -1}, \\
  S_4 = \sum\limits_{m = 1}^{\infty} (e^{-iqam} - e^{iqam})e^{ikam} = -i \frac{\sin qa}{\cos ka - \cos qa}
  \simeq  -\frac{2i}{ka}\cdot \frac{1}{n^2 -1}.
\end{eqnarray*}
By equating to zero the  determinant of the system
\eqref{eq:phi0}-\eqref{eq:phi1}, one obtains the dispersion
equation of the form:
\begin{equation}
  1 = \frac{2\pi i}{kb}  \left( F_0 + F_1 \right) S_3 + \left( \frac{2\pi i}{kb} \right)^2 F_0F_1
  \left( S_4^2 - S_3^2 \right).
  \label{eq:dips_anis}
\end{equation}
Simple (though cumbersome) arithmetic transforms of this equation
with substituted values of the sums $S_3$ and   $S_4$  as well as
the expressions for scattering amplitudes $F_0$ and  $F_1$
(formulas \eqref{eq:amp10_s_val}-\eqref{eq:amp1_s_val}) give the
final expression for the refractive index of a crystal
\begin{equation}
  n^2 \simeq 1 + \frac{4\pi}{k^2\Omega_2}\cdot
  \left\{ \frac{A_0}{1 + i\pi A_0} + \frac{A_1}{1 + i\frac{\pi}{2}A_1} \right\}.
  \label{eq:n2_anis}
\end{equation}
Note that  a three-dimensional case can be  considered in a
similar manner. As a result, if the amplitude of scattering by a
single center has the form $A(\theta)=A_0+A_1\cos\theta$, then the
refractive index of the crystal can be written as follows:
\begin{equation}
  n^2 = 1 + \frac{4\pi}{k^2 \Omega_3} \left\{  \frac{A_0}{1 + ikA_0}
  + \frac{A_1}{1+i\frac{k}{3}A_1}\right\},
  \label{eq:3D}
\end{equation}
where $\Omega_3$ is the volume of the crystal unit cell.

Under the assumption of smallness of the scattering amplitude and
with the above-selected form of its angular dependence, the
obtained expressions hold  for any values of $k$ far from the
diffraction conditions.
Moreover, the scatterers can obviously be arbitrary, not
necessarily  threads.
In a particular case of  crystals built from metallic threads,
substitution  into \eqref{eq:n2_anis} of  $A_0$ and  $A_1$  for
the wave with perpendicular polarization  (see formulas
\eqref{eq:c_perp}, \eqref{eq:asymptotic}) gives for the refractive
index the value $n^2=1+\dfrac{\pi R^2}{\Omega_2} > 1$, which means
that in such crystals the Vavilov-Cherenkov effect can be observed
\cite{Gurnevich2009, Gurnevich2010}.

In quantum mechanics, the scattering process is usually described
using  the transition matrix $\mathbf T$ (which is non-Hermitian
in general).
   Recall that the diagonal element of this matrix is proportional to
   the amplitude of forward scattering. Equation \eqref{eq:3D}
   reflects the fact that for an ordered medium (crystal), in the expression for the refractive
   index,  the diagonal  element of the matrix $\mathbf T$ must be replaced by the diagonal element of the Hermitian
   reaction matrix $\mathbf K$ (the properties of the matrix $\mathbf K$ see in \cite{davydov58nucl}).

 A detailed analysis shows that in view of the above
results,  the  equations describing the dynamical diffraction of
waves in crystals  must be modified. Their general form, of
course, does not change:
 \begin{eqnarray}
\left( 1 - \frac{k^2}{k_0^2} \right) \varphi(\vec{k})
 + \sum\limits_{\vec{\tau}} g(\vec{\tau})
 \varphi(\vec{k}-\vec{\tau}) =  0,
 \label{eq:dynamical_system1}
 \\
 \Psi(\vec{r})  =  \sum\limits_{\vec{\tau}} \varphi(\vec{k}
 + \vec{\tau}) e^{i(\vec{k}+\vec{\tau})\vec{r}},
 \label{eq:dynamical_system2}
 \end{eqnarray}
 where $g(\vec{\tau})$ is  the structure amplitude,
 $\vec{\tau}$ is the reciprocal lattice vector of the crystal.
 However, in the expression for the structure amplitude, the  amplitude of scattering by a single thread is replaced by
 the effective amplitude of wave scattering by a thread
 in the  crystal:
 \begin{equation}
  g(\vec{\tau}) = \dfrac{4\pi}{k_0^2 \Omega_2}
 \left( \frac{A_0}{1 + i\pi A_0} + \frac{A_1}{1
 + i\frac{\pi}{2}A_1} \cdot \frac{\vec{k}(\vec{k} + \vec{\tau})}{k^2}
  \right).
 \label{eq:structure_ampl}
 \end{equation}
This result agrees well with that given in
\cite{Baryshevsky1966,BVG95Nuclearoptics}  for the isotropic case.
According to \cite{BVG95Nuclearoptics}, when calculating  the
structure amplitude, one must exclude from the imaginary part of
the scattering amplitude the contribution to the total cross
section that comes from elastic coherent scattering.

\section*{Conclusion}
This paper considers the process of propagation of waves in
natural crystals and artificial crystals formed by anisotropically
scattering centers. The interaction of waves with individual
scatterers is described in terms of the scattering amplitude.
Special consideration is given to taking account of multiple
rescattering of the initial wave by the  centers in a crystal.
The obtained expression \eqref{eq:n2_anis}, which relates the
refractive index  of a crystal to the scattering amplitude,
differs from the known formula for the refractive index of
non-periodic media \eqref{eq:n2_3Dchaotic} and allows one to
correctly describe wave attenuation in crystals. In particular, if
scattering by a single center is elastic, the refractive index of
the crystal, calculated in accordance with \eqref{eq:n2_anis}, is
a real value, e.i., the attenuation is absent, whereas application
of a conventional formula  \eqref{eq:n2_3Dchaotic} in this case
leads to an erroneous conclusion that $n$ has a nonzero imaginary
part.

The equations derived in this paper, which  describe the field in
crystals,  coincide with  standard equations of the dynamical
diffraction  theory  in crystals \cite{Pinsker, Batterman1964,
Lamb}. However, in the expression for the structure amplitude
$g(\vec{\tau})$, the  amplitude of scattering by a single center
must be  replaced by  the effective amplitude of wave scattering
by a center located in the crystal (which is described by the
Hermitian reaction matrix $\mathbf K$). This enables one to
correctly describe the effect related to  the attenuation of
coherent waves in crystals.

In the present paper, a detailed consideration of the
electromagnetic wave  refraction in a two-dimensional photonic
crystal built from parallel metallic thread is given by way of
example. An interesting result for such a crystal is that the
index of refraction for a wave with $\vec{E}$ polarized
perpendicular to the threads is greater than $1$, and the
Vavilov-Cherenkov effect can be observed in the crystal
\cite{Gurnevich2009, Gurnevich2010}. A general approach applied
here to the description of scattering enables one to obtain the
results that are valid for a wide range of cases without being
restricted to either electromagnetic waves or crystals built from
threads.  They can be of interest, in particular, for  studying
diffraction of cold neutrons in crystals, investigating of various
nanocrystalline materials, designing metamaterials with prescribed
properties, etc.

\section*{References}

\begin{thebibliography}{10}

\bibitem{metamaterials}
S.~Zouhdi, A.~Sihvola, and A.P. Vinogradov.
\newblock {\em Metamaterials and Plasmonics: Fundamentals, Modelling,
  Applications}.
\newblock NATO Science for Peace and Security Series B: Physics and Biophysics.
  Springer, 2008.

\bibitem{Nanorods}
M\'ario~G. Silveirinha, Pavel~A. Belov, and Constantin~R.
Simovski.
\newblock Subwavelength imaging at infrared frequencies using an array of
  metallic nanorods.
\newblock {\em Phys. Rev. B}, 75:035108, Jan 2007.

\bibitem{Baryshevsky}
V.G. Baryshevsky and A.A. Gurinovich.
\newblock Spontaneous and induced parametric and Smith-Purcell radiation from
  electrons moving in a photonic crystal built from the metallic threads.
\newblock {\em Nuclear Inst. and Meth. B}, 252(1):92 -- 101, 2006.

\bibitem{FirstGridExp}
V.G. Baryshevsky, K.G. Batrakov, N.A. Belous, A.A. Gurinovich,
A.S. Lobko, P.V.,   Molchanov, P.F. Sofronov, and V.I. Stolyarsky.
\newblock {First observation of generation in the backward wave oscillator with
  a "grid" diffraction grating and lasing of the volume FEL with a "grid"
  volume resonator}
\newblock {\em LANL e-print ArXiv: physics/0409125}.

\bibitem{FirstGridExp1}
V.G. Baryshevsky,  N.A. Belous, A.A. Gurinovich, A.S. Lobko, P.V.,
Molchanov,  and V.I. Stolyarsky.
\newblock {Experimental study of a volume free electron laser with a 'grid'
resonator}
\newblock {\em Proc. of FEL 2006 BESSY, Berlin, Germany}, pages
331--334.
\newblock {\em LANL e-print ArXiv: physics/0605122}.



\bibitem{Goldberger}
M.L. Goldberger and K.M. Watson.
\newblock {\em Collision Theory}.
\newblock Structure of matter series. Wiley, 1975.

\bibitem{Baryshevsky1966}
V.G. Baryshevsky.
\newblock {\em Sov. Journal of Experimental and Theoretical Physics},
  51:1587--1591, 1966.

\bibitem{BVG95Nuclearoptics}
V.G. Baryshevsky.
\newblock {\em High-Energy Nuclear Optics of Polarized Particles}.
\newblock World Scientific, Singapore, 2012.

\bibitem{Gurnevich2009}
V.G. Baryshevsky and E.A. Gurnevich.
\newblock {The possibility of Cherenkov radiation generation in a photonic
  crystal formed by parallel metallic threads}.
\newblock {\em Vestnik BSU (The Journal of the Belarusian State University),
  ser.~1, No.3}, (3):38--44, 2009.

\bibitem{Gurnevich2010}
V.G. Baryshevsky and E.A. Gurnevich.
\newblock The possibility of Cherenkov radiation generation in a photonic
  crystal formed by parallel metallic threads.
\newblock {\em Proc. of 2010 International Kharkov Symposium on Physics and
  Engineering of Microwaves, Milimeter and Submilimeter Waves (MSMW-2010),
  21-26 June 2010}, pages 1--3.

\bibitem{Nikolsky}
V.V. Nikolsky and T.I. Nikolskaya.
\newblock {\em Electrodynamics and Radio Waves Propagation [in Russian]}.
\newblock Moscow, Nauka, 1989.

\bibitem{landau77quant}
L.D. Landau and E.M. Lifshitz.
\newblock {\em Quantum Mechanics: Non-Relativistic Theory}.
\newblock Statistical Physics. Pergamon Press, 1977.


\bibitem{Janke}
E.~Jahnke, F.~Emde, and F.~L{\"o}sch.
\newblock {\em Tables of Higher Functions}.
\newblock B. G. Teubner, 1966.

\bibitem{Belov1}
P~A Belov, S~A Tretyakov, and A~J Viitanen.
\newblock Dispersion and reflection properties of artificial media formed by
  regular lattices of ideally conducting wires.
\newblock {\em J. of Electromagn. Waves and Appl.}, 16(8):1153--1170, 2002.

\bibitem{davydov58nucl}
A.S. Davydov.
\newblock {\em Theory of the atomic nucleus [in Russian]}.
\newblock Fizmatgiz, Moscow, 1958.

\bibitem{Pinsker}
Z.G. Pinsker.
\newblock {\em Dynamical Scattering of X-Rays in Crystals}.
\newblock Springer Series in Solid-State Sciences. Springer-Verlag, 1978.


\bibitem{Batterman1964}
B.~Batterman and H.~Cole.
\newblock Dynamical diffraction of x rays by perfect crystals.
\newblock {\em Reviews of Modern Physics}, 36(3):681--717, 1964.

\bibitem{Lamb}
Willis~E. Lamb.
\newblock Capture of neutrons by atoms in a crystal.
\newblock {\em Phys. Rev.}, 55:190--197, Jan 1939.

\end{thebibliography}


\end{document}